# Exploration of Cryptocurrency Mining-Specific GPUs in AI Applications: A Case Study of CMP 170HX

XING KANGWEI

xkw@stmail.ujs.edu.cn

## Abstract

Since Ethereum's transition to the Proof-of-Stake (PoS) mechanism in 2023, NVIDIA's cryptocurrency mining-specific GPUs (e.g., the CMP series), designed with performance reductions in general computing and graphics rendering capabilities, have faced severe limitations in general-purpose computing scenarios, becoming typical electronic waste. This study systematically tests a computational power reuse scheme proposed by the open-source community—disabling specific instruction sets (e.g., Fused Multiply-Add instructions) through CUDA source code modifications—on the NVIDIA CMP 170HX platform. Experimental results validate the effectiveness of this approach, partially restoring the GPU's computational capabilities in artificial intelligence (AI) tasks. Performance evaluations using open-source GPU benchmarks (OpenCL benchmark, mixbench) and AI benchmarks (LLAMA-benchmark) reveal that its FP32 floating-point performance exceeds 15 times the original capability, while inference performance for certain precision levels in large language models surpasses threefold improvements. Furthermore, based on hardware architecture analysis, this paper proposes theoretical conjectures for further improving computational utilization through alternative adaptation pathways. Combining energy efficiency ratios and cost models, the recycling value of such obsolete GPUs in edge computing and lightweight AI inference scenarios is evaluated. The findings demonstrate that rationally reusing residual computational power from mining GPUs can significantly mitigate the environmental burden of electronic waste while offering cost-effective hardware solutions for low-budget computing scenarios.

# Graph

# Table

# 1. Introduction

## 1.1 Background

### *1.1.1 NVIDIA CMP GPUs and Their Sales Estimates*

In a blog post dated February 18, 2021 [1], NVIDIA stated, "GeForce is made for gaming, CMP is made for cryptocurrency mining," announcing the launch of NVIDIA CMP (Cryptocurrency Mining Processor), a product line specifically designed to meet the unique demands of Ethereum mining. The company emphasized that "CMPs do no graphics, have no display outputs, and feature lower peak core voltage and frequency, optimized for maximum mining performance and efficiency. CMPs do not meet the specifications of GeForce GPUs, ensuring gamers retain access to GeForce products." Subsequently, four CMP models—30HX, 40HX, 50HX, and 90HX—were showcased on NVIDIA's CMP product page [2]. However, this was not NVIDIA's first venture into mining-specific GPUs. According to GPU DATABASE [3], NVIDIA initially released the P106-100, a GPU dedicated to cryptocurrency mining, as early as June 19, 2017, followed by subsequent models such as the P106-090, P104-100, and P102-100.

NVIDIA has only disclosed CMP sales in select fiscal reports [4]. Based on publicly available data, the company's cryptocurrency-related revenue for its four fiscal quarters in 2022 was 155million,155million,266 million, 105 million, and 105million,and24 million, respectively. NVIDIA has historically obscured revenue generated from cryptocurrency mining, a practice that led to penalties from the U.S. Securities and Exchange Commission (SEC) [5]. Consequently, the Manufacturer's Suggested Retail Price (MSRP) for these CMP GPUs cannot be ascertained through official channels. According to third-party investigations [6], the approximate market prices of these CMP cards are summarized in the table below:

**Table 1-1** Prices and Theoretical Performance of CMP GPUs in 2022

| Model | Release Date | 2021 Average Price (USD) | FP16 Computational Power (TFLOPS) |
|---|---|---|---|
| CMP 30HX | 2021 Q1 | ~700-800 | 10.05 |
| CMP 40HX | 2021 Q1 | ~600-700 | 15.21 |
| CMP 50HX | 2021 Q2 | ~700-900 | 22.15 |
| CMP 90HX | 2021 Q2 | ~1400-1700 | 21.89 |
| CMP 170HX | 2021 Q3 | ~4000-5000 | 50.53 |

Based on the midpoint pricing, sales estimates for each GPU under different scenarios are derived as follows:

**Table 1-2** Estimated Sales Volume of CMP GPUs

| Model | Estimated ASP ($) | Scenario A (15/25/25/20/15) | Scenario B (25/30/20/15/10) | Scenario C (10/15/20/25/30) |
|---|---|---|---|---|

|  |  | Estimated Sales | Estimated Sales | Estimated Sales |
|---|---|---|---|---|
| CMP 30HX | 750 | ~110,000 | ~183,333 | ~73,333 |
| CMP 40HX | 650 | ~211,538 | ~253,846 | ~126,923 |
| CMP 50HX | 800 | ~171,875 | ~137,500 | ~137,500 |
| CMP 90HX | 1550 | ~70,968 | ~53,226 | ~88,710 |
| CMP 170HX | 4500 | ~18,333 | ~12,222 | ~36,667 |
| Whole |  | ~582,714 | ~640,127 | ~463,133 |

It is evident that the data derived from this inference are significantly lower than actual figures (excluding non-2022 sales and non-CMP series P106/P104/P102 models). Nonetheless, it is clear that hundreds of thousands of mining-specific GPUs have been either discarded or redirected to secondary markets following Ethereum's transition to PoS.

*1.1.2 Ethereum Mining Principles*

Wang et al. (2020, Chapter 4, Section 4.1.1, pp. 88–91) [7] describe Ethereum's GPU-dependent Proof-of-Work (PoW) mining process. In this mechanism, ETH selects the block with the highest difficulty as the valid block. Miner nodes generate blocks, while other nodes verify them. Miners attempt to find a suitable Nonce value through rapid computation to produce a result below the specified difficulty threshold. This process relies heavily on brute-force enumeration, whereas verifying the Nonce's validity is straightforward.

The bottleneck of this algorithm lies in memory read/write performance, preventing miners from improving efficiency through faster hardware.

First, the network maintains a Directed Acyclic Graph (DAG) dataset occupying approximately 4GB of memory. In GPU mining, the DAG is loaded into VRAM and updated every 30,000 blocks. If the DAG is not pre-generated, mining nodes halt block production.

The Nonce-solving process proceeds as follows:

Step 1: Perform SHA-3 computation on the block header input and a hypothetical Nonce value to generate a 128-byte Mix0.
Step 2: Read page data from the DAG at a position linked to Mix0.
Step 3: Mix Mix0 with the retrieved DAG page data to generate a 128-byte Mix1.
Step 4: Repeat Steps 2 and 3 64 times, resulting in a 128-byte Mix64.
Step 5: Process Mix64 into a 32-byte data digest.

Step 6: If the digest is less than or equal to the difficulty threshold, the Nonce is valid; otherwise, iterate with the next Nonce.

*1.1.3 GPU Export Restrictions to China*

Since October 2018 [8], when the Trump administration severed ties between Chinese chipmaker Fujian Jinhua Integrated Circuit Company and U.S. suppliers, the U.S. government has imposed restrictions on semiconductor technology and high-performance GPU exports to mainland China. In August 2022 [9], NVIDIA disclosed in an SEC filing that it was prohibited from exporting A100 and H100 GPUs to mainland China. By October 2023 [10][11], as stated in two additional SEC reports, the U.S. government expanded the ban to include GPUs such as the 14090, A800, and H800. In April 2025 [12], NVIDIA announced further restrictions blocking exports of products like the H20 to mainland China. Similarly, AMD [14] reported in SEC filings that its MI308 GPUs would be barred from export to the region.

**1.2 Research Objectives**

Given the aforementioned context, the following conclusions emerge:

Massive quantities (exceeding hundreds of thousands) of mining-specific GPUs face obsolescence.
Ethereum's Proof-of-Work (PoW) mining process relies heavily on GPU memory bandwidth and partial computational capabilities.
Regions such as mainland China are subject to U.S.-imposed restrictions on high-performance GPU imports.
Consequently, this study aims to validate whether these discarded mining-specific GPUs retain viable floating-point computational capabilities and AI computing capacity (for inference or training), thereby partially alleviating regional computational shortages caused by U.S. export controls in areas like mainland China.

Specific objectives of this study include:

Testing the floating-point performance of the CMP 170HX mining-specific GPU across half-precision (FP16), single-precision (FP32), and double-precision (FP64) formats.
Evaluating the CMP 170HX's performance under different GPU computing frameworks (CUDA/OpenCL).
Assessing the AI inference capabilities of the CMP 170HX using the LLAMA-bench benchmark.
Measuring the energy efficiency ratio (token/W) of the CMP 170HX during AI inference tasks.

**1.3 Testing Tools and Benchmark Overview**

### 1.3.1 MixBench

MixBench is a benchmarking tool [14] designed to evaluate the performance limits of GPUs (or CPUs) under mixed operational intensity kernels. The kernels executed are tailored for varying operational intensity values. Modern GPUs can hide memory latency by switching to threads capable of executing computational operations. This tool evaluates the actual optimal balance between memory and compute operations for a device. It supports implementations in CUDA, HIP, OpenCL, and SYCL for GPUs, and OpenMP for CPUs.

The tool conducts four types of experiments combined with global memory access:
Single-precision floating-point operations (Fused Multiply-Add, FMAs).
Double-precision floating-point operations (FMAs).
Half-precision floating-point operations (FMAs, GPU-only).
Integer multiply-add operations.
MixBench incrementally increases computational pressure and reports:

Compute iters (number of computational iterations per task).
Flops/byte (floating-point operations per byte).
Ex.time (execution time).
GFLOPS (giga floating-point operations per second; for integer operations, Iops/byte).
GB/sec (memory bandwidth in gigabytes per second).

### 1.3.2 OpenCL-Benchmark

OpenCL-Benchmark[15] is a small OpenCL benchmarking tool designed to measure peak performance of GPUs and CPUs. It supports any GPU on Windows, Linux, macOS, and Android operating systems.
The tool evaluates computational performance for FP64 (scalar), FP32 (scalar), FP16 (half2), INT64 (scalar), INT32 (scalar), INT16 (short2), INT8 (dp4a), memory bandwidth (coalesced/misaligned read/write), and PCIe bandwidth (send/receive/bidirectional). Specifically, the performance metrics for single-precision floating-point (FP32), double-precision floating-point (FP64), and half-precision floating-point (FP16) are measured in TFLOPs (trillion floating-point operations per second). For 64-bit integers (INT64), 16-bit integers (INT16), and 8-bit integers (INT8), performance is quantified using TIOPs (trillion integer operations per second).

### 1.3.3 GPU-Burn

GPU-Burn is a command-line-based, open-source stress-testing tool designed to maximally impose computational load on GPUs (Graphics Processing Units) [28]. Its primary purpose is to evaluate the stability, thermal performance, and computational accuracy of GPUs under prolonged periods of intensive workloads.

The tool supports configuring parameters such as runtime duration, VRAM utilization, enabling/disabling Tensor Cores, and selecting computation modes (FP32, FP16, FP64).

In this study, GPU-Burn will serve as the control group without any modifications to its source code or compilation parameters, and will be directly compiled and used as-is.

### *1.3.4 PyTorch*

PyTorch[16] is an open-source deep learning framework developed by the Facebook Artificial Intelligence Research (FAIR) team. It serves as a tensor library optimized for deep learning, leveraging both GPUs and CPUs, and enables the construction of deep neural networks using a tape-based autograd system.

Typically, PyTorch is utilized for two primary purposes:
a. As a replacement for NumPy, harnessing the computational power of GPUs;
b. As a deep learning research platform that provides maximum flexibility and performance efficiency.

By leveraging PyTorch's GPU support, we can develop scripts to evaluate the floating-point performance of the nVidia CMP 170HX.

### *1.3.5 llama.cpp*

Llama.cpp[17] is a tool implemented in pure C/C++ for inference of Meta's LLaMA models (and other models). Its primary goal is to enable large language model inference with minimal setup and state-of-the-art performance on both local machines and in the cloud.

For this performance test, we will use **llama-bench[18]**, a performance testing tool for Llama.cpp. It allows configuration of parameters such as input prompt length, generated text length, batch size, and number of test repetitions. Users can also enable attention optimizations (e.g., Flash Attention) and specify offloading layers to the GPU. Additionally, test results can be exported in formats like Markdown, CSV, JSON, JSONL, or SQL for easier organization.

LLAMA-BENCH supports three test types:
1. Prompt Processing (pp): Batch process prompts (-p).
2. Text Generation (tg): Generate a sequence of tokens (-n).
3. Prompt Processing + Text Generation (pg): Process prompts then generate tokens (`-pg`).

All options can be specified multiple times to run multiple tests. Each `pp` and `tg` test executes all combinations of the specified options.

Each test can be repeated a specified number of times, with results averaged. Outputs include the average tokens per second (t/s) and standard deviation. Some formats (e.g., JSON) also retain individual results for each repetition.

## 2. Testing Environment and Hardware Specifications

### 2.1 Tested GPU Hardware Parameters

This experiment employs the NVIDIA CMP 170HX 8GB as the test subject. Since NVIDIA has not released a whitepaper for this product, specifications are derived from publicly available data on GPU DATABASE [20] and inferred using NVIDIA A100 documentation [21][22]. The NVIDIA A100 and CMP 170HX share the same core architecture, boost clock frequency, thermal design power (TDP), and similar memory speeds. By scaling the CUDA core count ratio between the A100 and CMP 170HX, the theoretical performance ceiling of the CMP 170HX can be estimated.

#### 2.1.1 General Specifications

**Table 2-1** General Specifications of the nVidia CMP 170HX GPU

| Parameter | Value | Parameter | Value |
|---|---|---|---|
|  |  |  |  |
| GPU Name | NVIDIA CMP 170HX | Microarchitecture | Ampere |
| GPU Codename | GA100-105F-A1 | Process Node | 7 nm (TSMC) |
| Transistor Count | 826 mm² | TDP | 250 W |
| Bus Interface | PCIe x4 1.1 |  |  |

#### 2.1.2 Core Configuration

**Table 2-2** Core Specifications of the nVidia CMP 170HX GPU

| Parameter | Value | Parameter | Value |
|---|---|---|---|
| Base Clock | 1140 MHz | Boost Clock | 1410 MHz |
| CUDA Cores | 4480 | TMUs | 280 |
| Streaming Multiprocessors | 70 | Tensor Cores | 280 |
| L1 Cache | 192 KB per SM | L2 Cache | 8 MB |

#### 2.1.3 Memory Specifications

**Table 2-3** Memory Specifications of the nVidia CMP 170HX GPU

| Parameter | Value | Parameter | Value |
|---|---|---|---|
| Memory Type | HBM2e | Memory Size | 8 GB |
| Memory Clock | 1458 MHz | Effective Memory Clock | 2916 MHz |
| Memory Bus Width | 4096-bit | Memory Bandwidth | 1493 GB/s |

*2.1.4 Theoretical Performance*

**Table 2-4** Theoretical Performance of the nVidia CMP 170HX GPU

| Parameter | Value | Parameter | Value |
|---|---|---|---|
| Texture Fill Rate (Base) | 319.2 GT/s | Boost Texture Fill Rate | 394.8 GT/s |
| Boost FP16 Performance | 50.53 TFLOPS | Boost FP32 Performance | 12.63 TFLOPS |
| Boost FP64 Performance | 6.317 TFLOPS | Ethereum Hash Rate (Boost) | 164 MH/s |

*2.1.5 Supported Features*

**Table 2-5** Feature Support of the nVidia CMP 170HX

| Parameter | Value | Parameter | Value |
|---|---|---|---|
| OpenCL | 3.0 | Vulkan | 1.3 |
| CUDA | 8.0 | | |

## 2.2 Testing Platform

Due to experimental constraints, the GPU was tested via an Oculink-connected PCIe interface on a custom-built mini-PC platform. Oculink (Optical Link), developed by the PCI-SIG, is a cable-based internal/external connectivity standard based on the PCI Express (PCIe) protocol [22][23]. It provides a more flexible PCIe transmission method compared to traditional PCIe slots, particularly suited for servers, storage systems, and external PCIe devices (e.g., external GPU enclosures). Oculink employs specialized connectors, typically SFF-8611 (socket) and SFF-8612 (plug).

*2.2.1 Test Platform Configuration*

The hardware and software specifications of the test platform are as follows:

**Table 2-6** Test Platform Configuration

| Parameter | Value | Parameter | Value |
|---|---|---|---|
| CPU | AMD Ryzen 7 7840HS | CPU Core Architecture/Codename | Zen4/Phoenix |
| CPU Core | 8C/16T | Base/Max Boost Clock | 3.8 GHz / 5.1 GHz |
| s/Threads | | | |
| CPU Instruction Sets | MMX, SSE, SSE2, SSE3, SSE4.1, SSE4.2, SSE4a, AES, AVX, AVX2, AVX-512, BMI1, BMI2, F16C, FMA3, BF16, SHA, ABM, AMD64 | CPU TDP | 54 W |
| Memory Bus Width | 128-bit | Memory Frequency | 5600 MT/s |
| Operating System | Debian GNU/Linux 12.9 | Kernel Version | linux-kernel 6.1.0-31-amd64 |
| NVIDIA Driver Version | nvidia-linux-driver-570.86.16 | CUDA Version | 12.8 |

*2.2.2 Test Tool Configuration*

All testing tools were compiled from source code retrieved from their respective GitHub repositories as of March 20, 2025. By default, compilation parameters remained unmodified except for adjustments to CMakeLists.txt files. For certain tools, source code modifications were necessary to disable FMA (Fused Multiply-Add) instructions.

a. Mixbench

Following the developer's compilation guidelines, an initial build was performed using default settings. To validate FMA disablement, CMakeLists.txt was modified as follows:

**Table 2-7** MixBench Configuration

| Configuration | Default | FMA Disabled |
|---|---|---|
| CMakeLists Modifications | None | set(CUDA_NVCC_FLAGS "${CUDA_NVCC_FLAGS} -fmad=false" CACHE STRING "nvcc flags" FORCE)string(APPEND CMAKE_CUDA_FLAGS " -fmad=false") |
| Compilation Commands | | mkdir buildcd buildcmake ../mixbench-cudacmake --build ./ |

b. OpenCL-Benchmark

The developer-provided compilation script was used to build and execute the tool. To assess FMA's impact on performance, the source code was modified as shown below:

Table 2-8 OpenCL-Benchmark Configuration

| Configuration | Default | FMA Disabled |
|---|---|---|
| Source Code Modifications | None | diff<br>diff --git a/src/lbm.cpp b/src/lbm.cpp<br>index d99202f..28aeb25 100644<br>--- a/src/lbm.cpp<br>+++ b/src/lbm.cpp<br>@@ -286,6 +286,8 @@ void LBM_Domain::enqueue_unvoxelize_mesh_on_device(const Mesh* mesh, const uchar<br> }<br> <br> string LBM_Domain::device_defines() const { return<br>+<br>+"\n #pragma OPENCL FP_CONTRACT OFF" // Disables implicit FMA optimizations<br>+"\n #define fma(a, b, c) ((a) * (b) + (c))" // Overrides OpenCL's fma() function<br> "\n #define def_Nx "+to_string(Nx)+"u"<br> "\n #define def_Ny "+to_string(Ny)+"u"<br> "\n #define def_Nz "+to_string(Nz)+"u" |
| Compilation Commands | ./OpenCL-Benchmark/make.sh | |

c. GPU-Burn

GPU-Burn served as a control group for floating-point performance testing. No modifications were made to its source code or compilation parameters.

Table 2-9 GPU-Burn Configuration

| Compilation Commands: |
|---|
| make |
| ./gpu_burn -tc 3600 |

d. PyTorch

A custom Python script based on PyTorch was developed to evaluate GPU floating-point performance through large-scale matrix operations. The script will be open-sourced on

GitHub.

e. llama.cpp

To ensure compatibility with the CMP 170HX's CUDA compute capability (8.0), modifications were made to ggml-cuda/CMakeLists.txt to disable FMA and enforce architectural optimizations:

**Table 2-10** llama.cpp Configuration

| Configuration | Default | FMA Disabled |
|---|---|---|
| CMakeLists Modifications | None | set(CUDA_FLAGS --fmad=false) |
| Compilation Commands | | cmake -B build -A x64 -DGGML_CUDA=ON -DCMAKE_CUDA_ARCHITECTURES="80"cmake --build build --config Release --parallel 16 |

## 3. Computational Performance Testing

Results of performance tests may exhibit deviations due to experimental constraints, and limitations persist in tools such as OpenCL-benchmark and Mixbench.

### 3.1 Single-Precision (FP32) Performance

FP32 , or Single-Precision Floating-Point , uses 32 bits to represent a floating-point number. Typically, this includes 1 sign bit, 8 exponent bits, and 23 mantissa bits, adhering to the IEEE 754 standard . It remains the most common and foundational floating-point format in GPU computing, offering a balanced precision and range. The peak floating-point performance of GPUs is generally measured in FP32 throughput (measured in TFLOPS , or trillion floating-point operations per second).

FP32 strikes a moderate precision level with broad applicability, serving as the standard precision for the majority of graphics rendering and general-purpose GPU (GPGPU) applications. It benefits from widespread hardware support and robust optimization. Rendering calculations in most games, professional visualization tools, and CAD/CAM applications rely on FP32 for processing vertex coordinates, texture coordinates, color values, lighting computations, and more. FP32 also forms the basis for many scientific computations, physics simulations, signal processing, and financial modeling tasks.

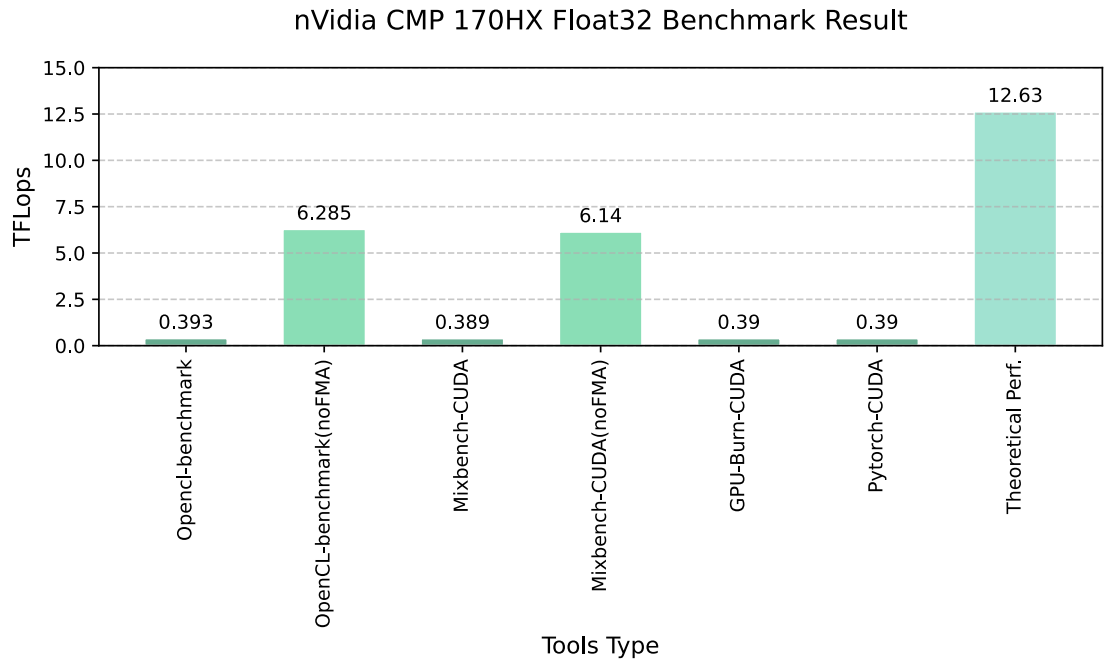

**Graph 3-1** nVidia CMP 170HX Float32 Benchmark Resault

By default , the FP32 performance of the 170HX is remarkably low, at approximately 0.39 TFLOPS , representing roughly 1/32 of its theoretical FP32 performance . This performance surpasses only the NVIDIA Tesla C870 (released in 2007), which had a theoretical FP32 throughput of about 0.346 TFLOPS [24].

However, when applying a suggestion from user niconini's blog post [25]—specifically, disabling Fused Multiply-Add (FMA) instructions during compilation—the CMP 170HX can achieve approximately 6.2 TFLOPS in benchmarks, reaching half of its theoretical FP32 performance . This performance surpasses that of the NVIDIA Tesla P6 [29].

### 3.2 Half-Precision (FP16) Performance

FP16 , or Half-Precision Floating-Point , uses 16 bits to represent a floating-point number. It typically includes 1 sign bit, 5 exponent bits, and 10 mantissa bits. Compared to FP32, FP16 has a narrower numerical range and lower precision. However, it offers faster computation speed (theoretically twice as fast as FP32) and reduced memory usage and bandwidth (half of FP32's requirements).

FP16 is primarily applied in deep learning training (specifically mixed-precision training ) and deep learning inference .

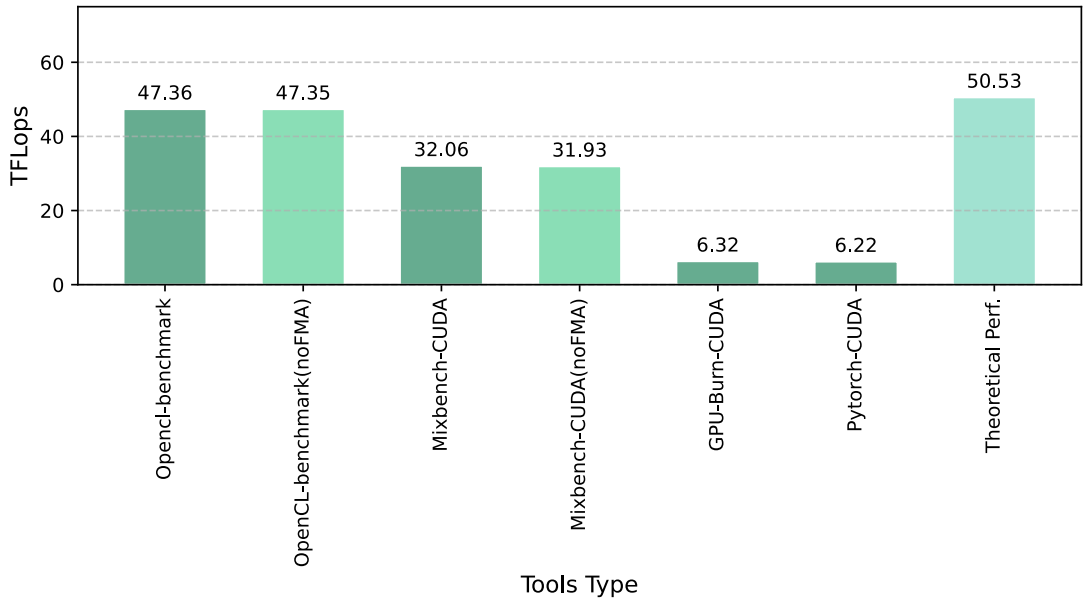

**Graph 3-2** nVidia CMP 170HX Float16 Benchmark Resault

The FP16 performance of the nVidia CMP 170HX remains unaffected regardless of FMA status (whether Fused Multiply-Add instructions are enabled or disabled). This performance level is approximately equivalent to that of the NVIDIA GeForce RTX 4080 desktop GPU's FP16 performance [27] (excluding contributions from Tensor Cores). Theoretically, this could enable the CMP 170HX to handle certain inference tasks that are not memory bandwidth-bound, such as inference for Stable Diffusion.

However, the FP16 performance reported by PyTorch and GPU-Burn is only around 6.3 TFLOPS, which is significantly lower than the results from OpenCL-Benchmark and Mixbench. The author speculates that this discrepancy likely arises from differences in how FP16 precision data is handled by PyTorch and GPU-Burn compared to the other tools.

Similar to the FP32 scenario, OpenCL-based benchmarks show slightly higher performance than CUDA-based ones**. This phenomenon may stem from the fact that the CUDA-based Mixbench benchmark(with 1,024 compute iterations and a Flops/Byte ratio of 512.250) does not fully stress the GPU, thereby failing to unlock its maximum potential.

### 3.3 Double-Precision (FP64) Performance

Double Precision Floating Point (FP64) uses 64 bits to represent a floating-point number. It typically consists of 1 sign bit, 11 exponent bits, and 52 mantissa bits, adhering to the IEEE 754 standard. FP64 offers extremely high numerical precision and an extensive range of representable values. Its exceptional precision effectively mitigates error accumulation, making it indispensable for calculations requiring rigorous numerical accuracy.

Scientific computing and engineering simulations are the core application domains of FP64. In fields demanding complex mathematical computations with highly precise results, such as computational fluid dynamics (CFD), finite element analysis (FEA), climate and meteorological modeling, astrophysics, molecular dynamics, and quantum chemistry, FP64 is an essential requirement.

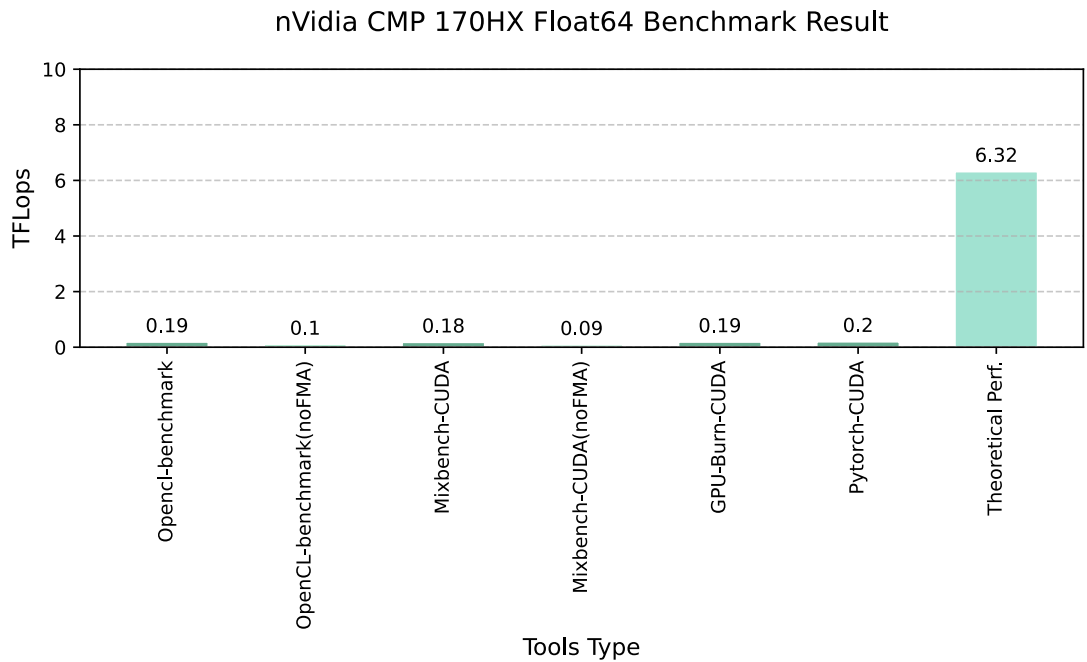

**Graph 3-3** nVidia CMP 170HX Float64 Benchmark Resault

The FP64 performance of the CMP 170HX is approximately 1/64 of its theoretical peak performance, and this drops to 1/128 when FMA (Fused Multiply-Add) is disabled. This performance is far behind that of mainstream scientific computing GPUs like the A100, rendering it barely capable of meeting the requirements of any scientific computing tasks.

**3.4 32-Bit Integer (INT32) Performance**

INT32 refers to a 32-bit integer. It can represent integers ranging from -2,147,483,647 to 2,147,483,647 (signed) or 0 to 4,294,967,295 (unsigned).
Integer operations are generally faster than floating-point operations and incur no precision loss within their representable range. They are commonly used for address calculations, indexing, bitmasking, shifting, logical operations (AND, OR, XOR), and other bit-level manipulations. INT32 is frequently employed in graphics rendering for state management, data packing/unpacking, and bitwise operations in specific algorithms (e.g., hashing).

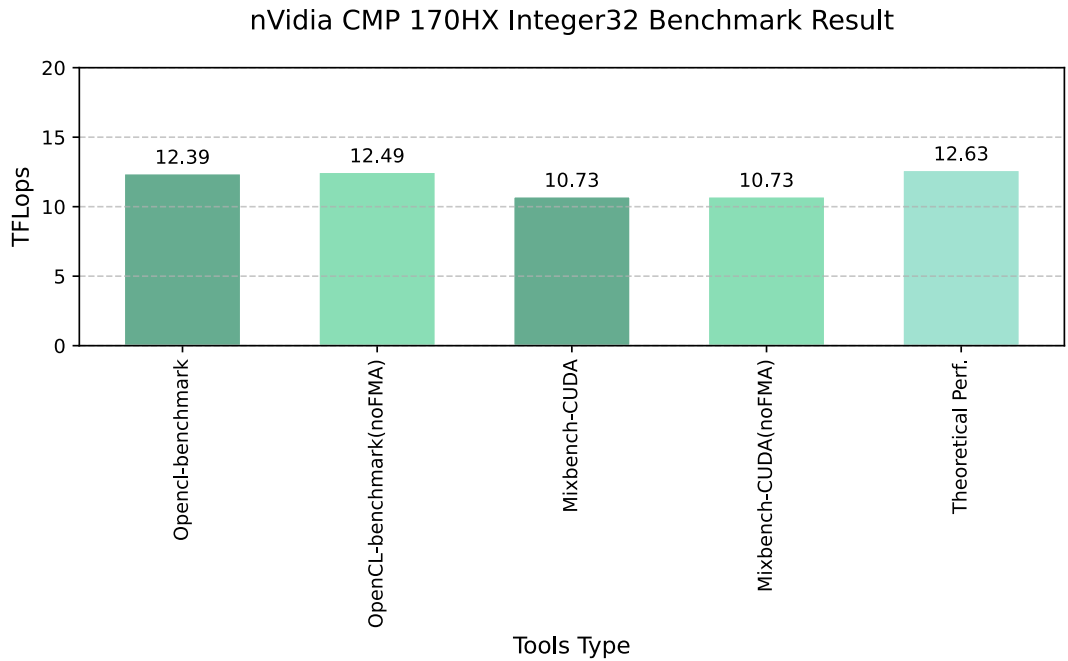

**Graph 3-4** nVidia CMP 170HX Integer32 Benchmark Resault

The INT32 performance of the NVIDIA CMP 170HX is not significantly restricted. However, applications relying heavily on this precision are relatively rare.

Similar to FP32 and FP16, and even more noticeably, the CMP 170HX's INT32 performance in OpenCL is slightly higher than in CUDA. This discrepancy likely stems from the CUDA-based mixbench benchmark tool's computational pressure (1024 Compute iters, Flops/byte = 512.250), which may not fully leverage the GPU's potential.

### 3.5 Memory Bandwidth Testing

GPU memory bandwidth is one of the key performance bottlenecks. Even if a GPU has strong computational power (high TFLOPS), if data cannot be delivered to compute units or retrieved from them in a timely manner, the compute units will idle and wait, resulting in actual performance far below the theoretical peak. Memory bandwidth is critical for all data-intensive tasks. During the decoding phase of LLMs (large language models), memory bandwidth becomes the primary bottleneck. For bandwidth-sensitive algorithms like Ethash, their performance is almost entirely determined by memory bandwidth.

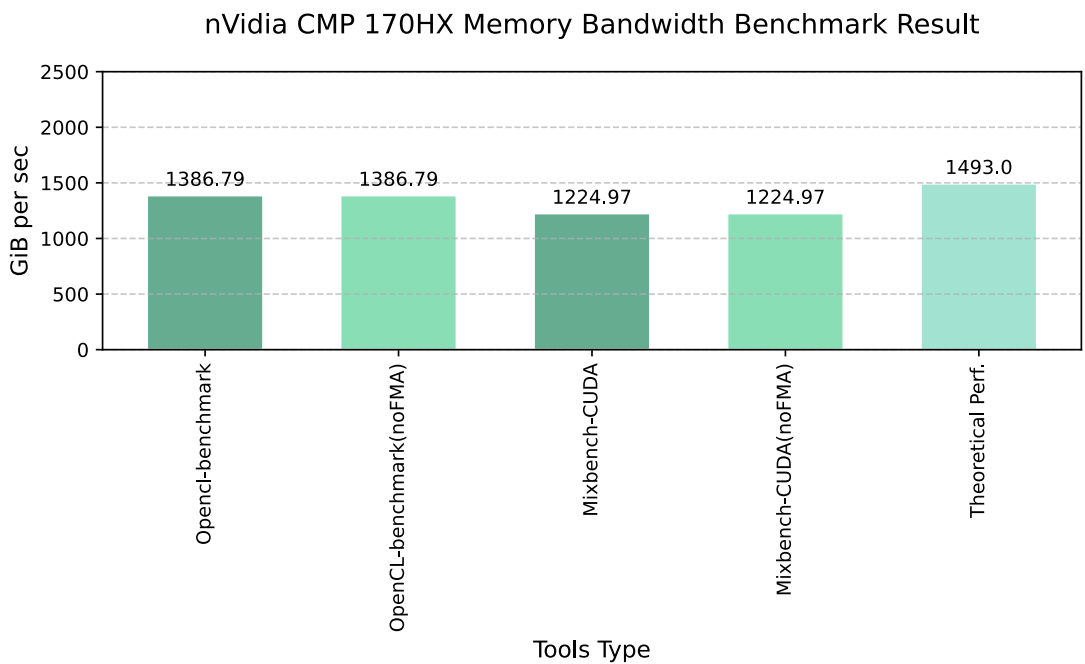

**Graph 3-5** nVidia CMP 170HX Memory Bandwidth Benchmark Resault

Since the Ethereum mining algorithm relies heavily on bandwidth [7], the memory bandwidth of the CMP 170HX is fully retained. However, compared to the standard NVIDIA A100/A800 GPUs with 40GB/80GB memory, its 8GB capacity renders it incapable of handling large-scale model loading.

Nevertheless, even with only 8GB of memory, the CMP 170HX can still gain some performance advantages in bandwidth-dependent applications (e.g., the decoding phase of large language models) due to its retained bandwidth efficiency.

## 4. Large Language Model Inference Performance Testing

### 4.1 Test Configuration

Considering the nVidia CMP 170HX has only 8GB HBM2e memory, we selected the Qwen2.5-1.5B model to ensure all layers of LLM weights could be loaded into GPU VRAM. We tested different quantization versions of this model under the ggml framework (f32, f16, q8_0, q6_k, q4_k_m, q2_k_m) across three scenarios: full model loading, prefilling 512 tokens, and decoding 128 tokens.

The 1.5B Qwen2.5[26] model is a transformer-based architecture developed by the Qwen team featuring RoPE, SwiGLU, RMSNorm, Attention QKV bias, and tied embeddings. It has a total of 1.54B parameters (1.31B excluding the embedding layer). The architecture includes 28 layers with 12 Q heads and 2 KV heads (GQA). The maximum context length supports 32,768 tokens.

Following the usage guidelines of llama-bench in llama.cpp[18], the testing command is:
./llama-bench -m ./models/Qwen2.5-1.5B-f32.gguf -ngl 28
Where:

## 4.2 Prefill Speed

The prefill stage involves the model processing input prompts. During this phase, all tokens in the input prompt are processed in parallel. For each token in the prompt, the model calculates corresponding Query (Q), Key (K), and Value (V) vectors, and uses self-attention mechanisms to compute inter-token relationships. The model computes and stores K and V vectors for all tokens across each Transformer block layer, forming cached contextual memory.

Due to parallel processing of the entire input sequence, this stage is computationally intensive (compute-bound) especially with long prompts. After controlling for prompt length, model size, and batch size, hardware compute capability determines performance. Theoretical performance estimation is calculated as:

$$u_d = \frac{u_o}{o_{sm}} \cdot d_{sm}$$

Where $u_d$ represents CMP 170HX's theoretical performance, $u_o$ is A100's measured performance, $o_{sm}$ is A100's SM count (108), and $d_{sm}$ is CMP 170HX's SM count (70).

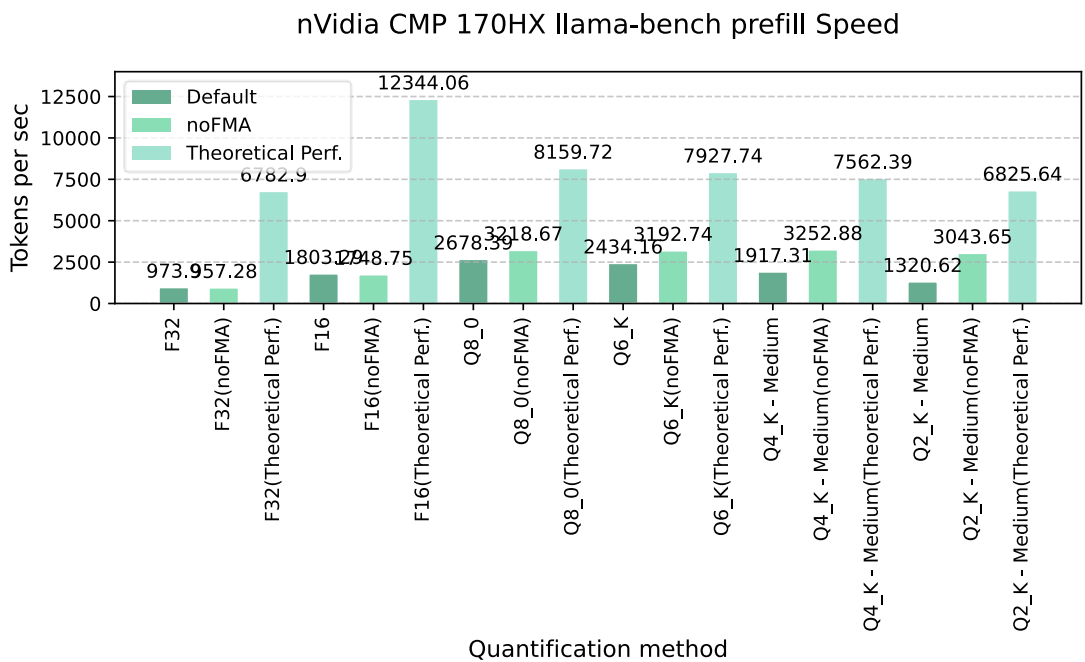

**Graph 4-1** nVidia CMP 170HX llama-bench prefill Speed

However, CMP 170HX's actual prefill performance lagged significantly behind theoretical expectations. Even with FMA disabled, prefill speeds only reached 14-45% of theoretical limits. This discrepancy likely stems from the inability to utilize Tensor Cores for acceleration. Disabling FMA significantly boosted prefill speeds for quantized models (Q8/Q6_K/Q4_K_M/Q2_K_M), with Q2_K_M achieving the highest improvement (231% of original rate). In contrast, f32/f16 models showed no performance gains.

**4.3 Decoding Speed**

After the prefill stage completes and the model generates the initial KV cache from the input prompt, the decoding phase begins. This stage aims to sequentially generate output tokens to form the final response or continuation.

The decoding process is autoregressive:

The model uses the hidden state of the last input token from the prefill stage to predict the first output token.
After generating the first output token, the model calculates its corresponding Q, K, V vectors.
The new token's Q vector computes attention with all previous tokens' K and V vectors (stored in the KV cache), including both the original prompt and previously generated tokens.
The new token's K and V vectors are added to the KV cache.
The next output token is predicted based on the attention results.
Steps 2-5 repeat until the end-of-text token (e.g., <|endoftext|>) is generated or the maximum output length is reached.
The decoding stage heavily depends on the KV cache. Each new token requires reading the full, growing KV cache from memory for attention calculations. While each step processes only one new token (resulting in lower computational load per step compared to prefill), the process is serial (token N cannot be generated until token N-1 is completed). Additionally, repeatedly accessing the large, linearly growing KV cache (its size increases with generated tokens) makes this stage memory-bound (bandwidth-limited).

After controlling for KV cache size, generated token count, model scale, and decoding strategies, memory bandwidth becomes the primary performance bottleneck. Theoretical performance estimation is calculated as:

$$u_d = \frac{u_o}{o_{bw}} \cdot d_{bw}$$

$u_d$ represents CMP 170HX's theoretical performance, $u_o$ is A100's measured performance, $o_{bw}$ is A100's memory bandwidth (1555 GB/s), and $d_{bw}$ is CMP 170HX's memory bandwidth (1493 GB/s).

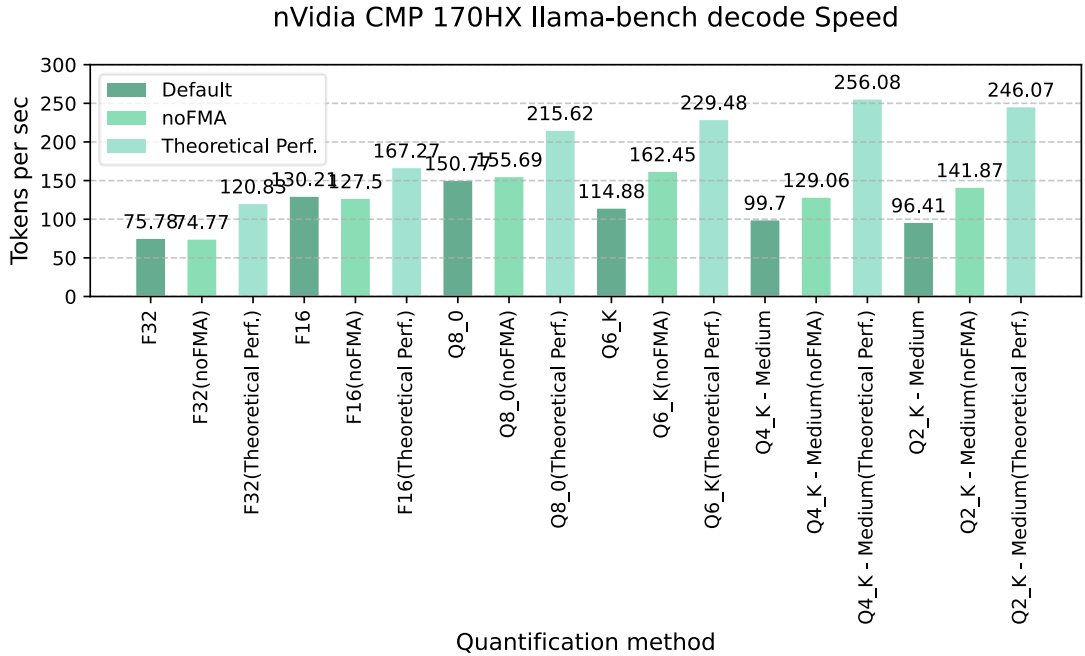

**Graph 4-2** nVidia CMP 170HX llama-bench decode Speed

Similar to the prefill stage, CMP 170HX's decoding speed also underperforms theoretical expectations. However, since memory bandwidth is not a bottleneck, it achieves 39-78% of theoretical performance. Disabling FMA further boosts decoding speeds for quantized models (Q8/Q6_K/Q4_K_M/Q2_K_M), reaching 50-78% of theoretical performance.

**4.4 Energy Efficiency Analysis**

Based on the charts above, we observe that the prefiling speed of large language models significantly exceeds their decoding speed. According to the data presented, the decoding phase dominates the inference latency in large language models. To simplify calculations, we directly use decoding speed per watt as the unit for energy efficiency measurement.

Power consumption was recorded using the NVIDIA-SMI tool during the inference process.

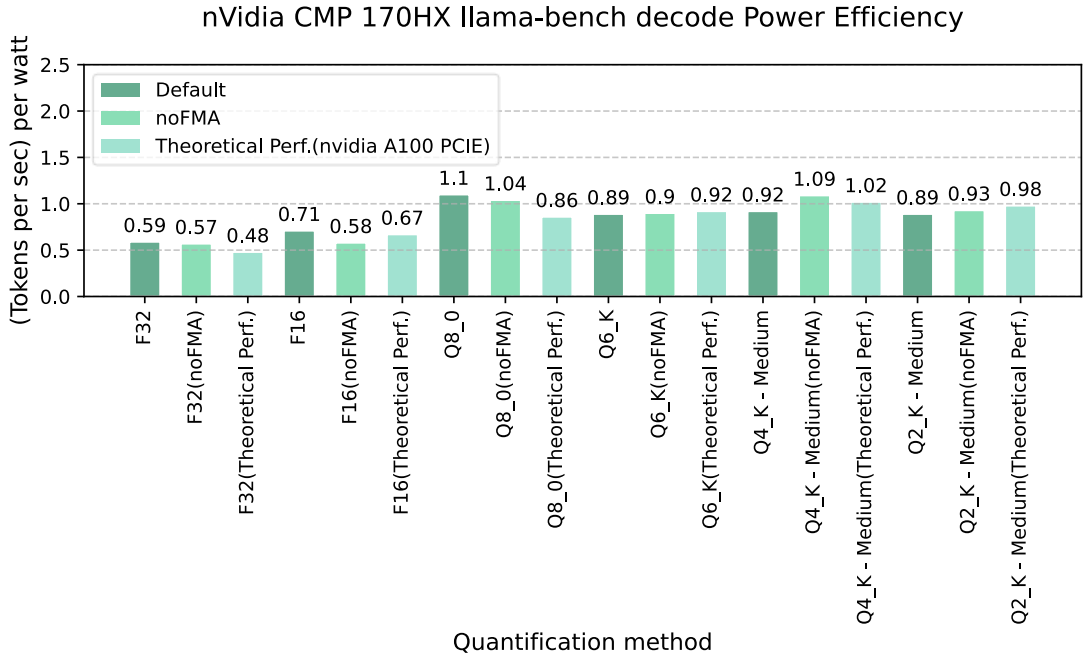

**Graph 4-3** nVidia CMP 170HX llama-bench decode Power Efficiency

In the charts, the energy efficiency of the CMP 170HX outperforms its theoretical efficiency (comparable to the A100) in half of the scenarios (F32, F16, Q8). Disabling FMA (Fused Multiply-Add) improves decoding speed for quantized large language models at Q6, Q4_K_M, and Q2_K_M levels on the CMP 170HX during inference. However, this also leads to a decline in energy efficiency.

Overall, the NVIDIA CMP 170HX demonstrates energy efficiency levels consistent with those of the GA100 core architecture when deployed for large language model inference.

## 5. Comprehensive Analysis and Discussion

The experimental conclusions presented above demonstrate the following:
a．The floating-point performance of the NVIDIA CMP 170HX can be leveraged through specific methods.
b．The memory bandwidth of the CMP 170HX grants it an advantage in large language model (LLM) inference.
c．Under full utilization of its performance, the energy efficiency of the CMP 170HX matches that of the A100 with the same GA100 core architecture.

### 5.1 Analysis of Floating-Point Performance

NVIDIA has imposed strict limitations on the floating-point performance of the CMP 170HX, primarily affecting FP32 and FP64 operations.

FP32 : Performance can be restored to approximately 6 TFLOPS by disabling Fused Multiply-Add (FMA) during CUDA compilation.

FP64 : No known method exists to recover its performance.

FP16 : Without Tensor Cores, FP16 appears unrestricted and can theoretically reach ~50 TFLOPS . However, frameworks like PyTorch fail to utilize this capability due to their inherent FP16 computation methods.

**5.2 Evaluation of AI Performance**

The CMP 170HX achieves energy efficiency comparable to the A100 only when the bottleneck of an application (e.g., LLM inference) is memory bandwidth . Disabling FMA improves decoding speed for quantized models (e.g., Q6, Q4_K_M) but reduces energy efficiency. Notably, disabling FMA enhances FP32 performance, suggesting that low-precision quantized models in Llama.cpp may rely heavily on FP32. The CMP 170HX may thus be suitable for highly quantized model inference.

Additionally, its integer performance remains uncrippled, enabling potential use for pure integer-based inference tasks.

**5.3 Current Limitations**

The widely used deep learning framework PyTorch cannot exploit the CMP 170HX's FP16 performance . While disabling FMA improves FP32 performance, efforts to modify PyTorch (e.g., altering static FMA code) face significant challenges:

PyTorch's massive codebase complicates manual adjustments to its CUDA-level FMA dependencies.

Unlike tools like MixBench, PyTorch's FP16 limitations may stem from deeper architectural constraints, requiring even more extensive modifications to its CUDA source code.

No viable solution exists for restoring FP64 performance.

**5.4 Theoretical Pathways for Full Performance Recovery**

The following methods could theoretically unlock the CMP 170HX's full computational potential, though each poses significant challenges:

a. Cracking NVIDIA drivers : Legally risky and technically demanding due to hardware-level restrictions.

b. Open-source driver approaches : Using NVIDIA's open-source kernel drivers and Vulkan-based user-space drivers might bypass closed-source limitations. However, stability, functionality, and compatibility issues persist, and restrictions may be embedded in the GPU System Processor (GSP).

c. Custom CUDA programming : Developing standalone CUDA programs (instead of relying on PyTorch) could fully utilize the CMP 170HX's 50 TFLOPS FP16 and 6 TFLOPS FP32 performance by avoiding FMA in code/compiler settings. This approach avoids framework limitations but requires manual optimization.

## 6. Conclusions and Recommendations

### 6.1 Key Conclusions

The nVidia CMP 170HX is an Ethereum-mining specialized GPU with significantly reduced floating-point computing performance but retained massive memory bandwidth. Manually writing CUDA code to disable FMA optimization can partially bypass its computational limitations (FP32). When running applications heavily reliant on memory bandwidth with memory requirements below 8GB, the CMP 170HX demonstrates energy efficiency comparable to the A100.

### 6.2 Application Recommendations

Thanks to its energy efficiency, the nVidia CMP 170HX is most suitable for use in regions lacking access to datacenter-level GPUs, such as community edge nodes that prioritize cost and service latency and security while not requiring advanced computational power. It can be employed for inference of small-scale large language models or highly memory bandwidth-intensive applications in cost-sensitive industrial simulations (e.g., FluidX3D).

Using the CMP 170HX requires knowledge of CUDA programming and GPU-related expertise. Currently, it is not feasible for gaming purposes.

## Acknowledgments

As an independent researcher, this research process encountered numerous setbacks. I am grateful to the predecessors in the Chinese internet community for guiding me toward a viable path. I thank developers worldwide for their contributions to open-source tools. Special thanks to Yuki (祈洱) for compiling and installing the Linux driver for the CMP 170HX.

As a visually impaired individual, I found myself caught in the economic downturn and trade wars following the pandemic after graduation. I have struggled to find suitable employment. I am deeply grateful to my parents for their care during this anxious period and their support for my passions. Without their encouragement, I might have chosen a dark and uncertain path long ago.

# Appendix

## Ex.1. Sales Estimation Methodology

Since NVIDIA's financial reports do not directly provide sales data for specific CMP card models, we need to estimate their sales volumes by analyzing total revenue figures and estimated average selling prices. The 2022 fiscal year marked the peak in CMP sales revenue, with total income reaching $550 million. We assume the majority of these sales occurred in the first three quarters of the 2022 fiscal year. Using the average selling price estimates from Table 2, we can make rough estimates of the sales volumes for each model. However, due to the lack of detailed sales structure data, we adopt a simplified approach based on total revenue and a weighted average price to calculate overall sales volume. More precise estimates would require granular quarterly revenue data and information about the sales mix proportions of different models across various periods. For instance, if the CMP 170HX contributed a larger portion of revenue during its initial high-price period, its actual sales volume might have been relatively lower compared to lower-priced models that likely sold in greater quantities.

## Ex.2. Additional Tests

### Ex.2.1. INT8 Performance

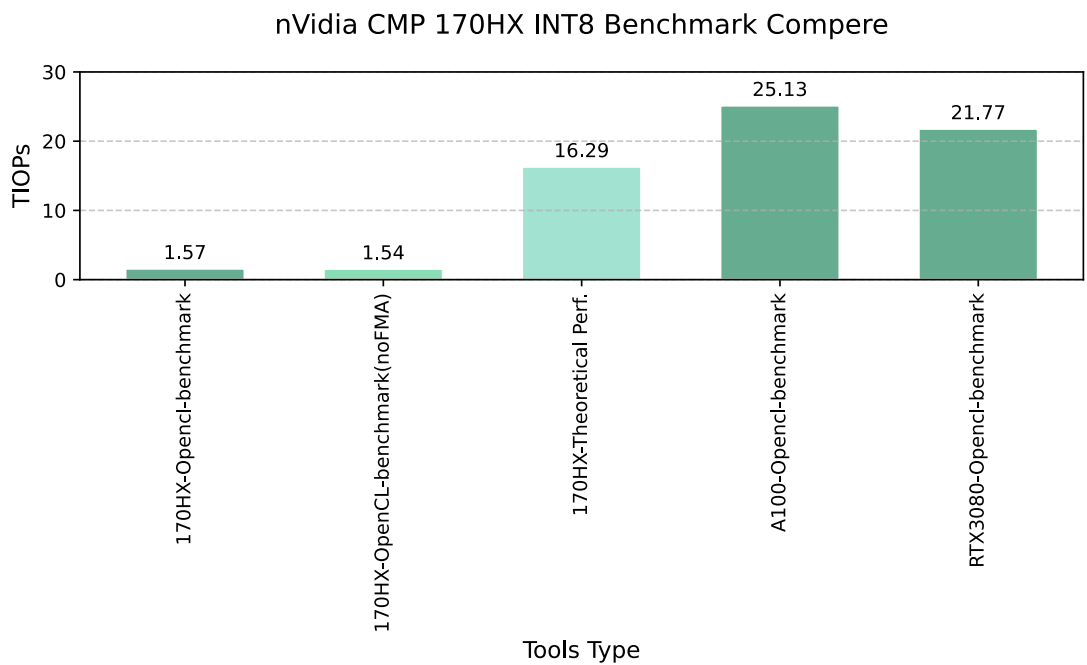

**Graph EX.1** nVidia CMP 170HX Integer8 Benchmark Resault

### Ex.2.2. PCIe Bandwidth Analysis

The CMP 170HX natively supports PCIe 1.1 x4 bandwidth. By populating additional coupling capacitors on the PCB, the interface can theoretically operate at PCIe 1.1 x16, though this modification was not validated in this study.

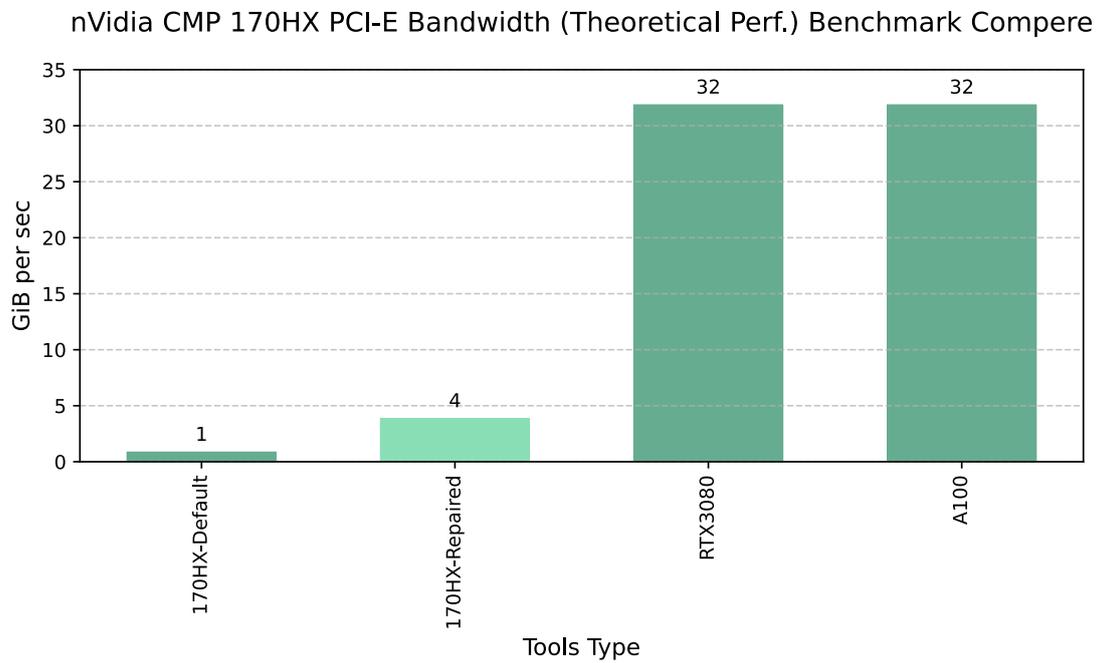

**Graph EX.2** nVidia CMP 170HX PCI-E Bandwidth Benchmark Resault